\documentclass[%
 reprint,
superscriptaddress,
 amsmath,amssymb,
 aps,
]{revtex4-1}
\usepackage[latin1]{inputenc}
\usepackage{graphicx}
\usepackage{dcolumn}
\usepackage{bm}
\usepackage{siunitx}
\usepackage[breaklinks]{hyperref}

\begin{document}

\title{Measurement of the in-plane thermal conductivity by steady-state infrared thermography} 

\author{Anton Greppmair}
\altaffiliation{Electronic mail: anton.greppmair@wsi.tum.de}
\affiliation{Walter Schottky Institut and Physik-Department, Technische Universit{\"a}t M{\"u}nchen, Am Coulombwall 4, 85748 Garching, Germany}

\author{Benedikt Stoib}
\affiliation{Walter Schottky Institut and Physik-Department, Technische Universit{\"a}t M{\"u}nchen, Am Coulombwall 4, 85748 Garching, Germany}

\author{Nitin Saxena}
\affiliation{Lehrstuhl f{\"u}r Funktionelle Materialien, Physik-Department, Technische Universit{\"a}t M{\"u}nchen, James-Franck-Strasse 1, 85748 Garching, Germany}

\author{Caroline Gerstberger}
\affiliation{Walter Schottky Institut and Physik-Department, Technische Universit{\"a}t M{\"u}nchen, Am Coulombwall 4, 85748 Garching, Germany}

\author{Peter M{\"u}ller-Buschbaum}
\affiliation{Lehrstuhl f{\"u}r Funktionelle Materialien, Physik-Department, Technische Universit{\"a}t M{\"u}nchen, James-Franck-Strasse 1, 85748 Garching, Germany}

\author{Martin Stutzmann}
\affiliation{Walter Schottky Institut and Physik-Department, Technische Universit{\"a}t M{\"u}nchen, Am Coulombwall 4, 85748 Garching, Germany}

\author{Martin S. Brandt}
\affiliation{Walter Schottky Institut and Physik-Department, Technische Universit{\"a}t M{\"u}nchen, Am Coulombwall 4, 85748 Garching, Germany}

\date{\today}

\begin{abstract}
We demonstrate a simple and quick method for the measurement of the in-plane thermal conductance of thin films via steady-state IR thermography. The films are suspended above a hole in an opaque substrate and heated by a homogeneous visible light source. The temperature distribution of the thin films is captured via infrared microscopy and fitted to the analytical expression obtained for the specific hole geometry in order to obtain the in-plane thermal conductivity. For thin films of poly(3,4-ethylenedioxythiophene):polystyrene sulfonate post-treated with ethylene glycol and of polyimide we find conductivities of \SI{1.0}{\watt\per\metre\per\kelvin} and \SI{0.4}{\watt\per\metre\per\kelvin} at room temperature, respectively. These results are in very good agreement with literature values, validating the method developed.
\end{abstract}


\maketitle

\section{Introduction: Thermal conductivity of thin films}
With a rising interest in the thermoelectric properties of meso-structured\cite{Isa14,Li13,Wan08,Dre07} and organic materials\cite{Sta12,Zhe14}, the full electric and thermal characterization of thin films becomes ever more relevant. Though the electrical parameters are often well studied, there is still little data on the thermal conductivity of novel thin film materials like organic semiconductors. Even for more standardized materials information about the dependence of the thermal conductivity on measurement orientation, material type, and film fabrication is scarce.\cite{Ben99,Yok95,Cho87} Therefore, a quick and versatile method to measure the thermal conductivity of films thinner than \SI{10}{\micro\meter} is needed, which is challenging for established methods.

Problematic for the measurement of the thermal conductivity of thin films can be the substrates often used to support the film under investigation. These substrates usually are orders of magnitude thicker than the film and therefore easily have a higher thermal conductance than the thin film itself. Similarly, heat transport through wires for local heaters or thermometers in contact with the films can disturb thermal conductivity measurements. Thus substrate- and contact-free methods are generally preferred. The Raman-shift method \cite{Bal11,Sto14b,Rep14,Sto14} is such a possible technique, but, while it has successfully been applied to well-understood materials such as SiGe alloys or graphene, the necessary complex calibration procedure somewhat restricts the use of this method for other materials. Furthermore, to ensure reasonable measurement durations and sensitivity, a high illumination power is typically applied in Raman experiments which generates high temperature differences and further limits this technique. Time-resolved optical pump-and-probe measurements\cite{Dal02} are an alternative, but also depend on calibration and smooth film surfaces without diffuse reflection. Transient 3$\omega$ on-chip measurements\cite{Koj15,Vol13} circumvent the need for calibration and are less restricted in terms of the type of film investigated, but sample preparation is very demanding. Since either sample mounting or measurement are time consuming in all of these techniques, they cannot easily provide information about the homogeneity of a film as a whole. 

A method which is quicker and allows easy on-sample calibration is IR thermography. For the determination of the in-plane thermal conductivity with thermography, the {\AA}ngstr{\o}m method, which is a dynamic measurement technique, is traditionally used.\cite{Wol06,Hor03} It, however, relies on well-defined illumination patterns generated, e.g., by focused laser beams. Here we demonstrate an IR thermography method for the measurement of the in-plane thermal conductivity of thin films where a steady-state temperature gradient is generated via homogeneous heating by strongly absorbed visible light and a substrate serving as a lateral heat sink. Below, we will introduce the measurement principle and present the necessary experimental setup and the calibration steps. We apply the method to thin films of poly(3,4-ethylenedioxythiophene):polystyrene sulfonate (\mbox{PEDOT:PSS}) and of polyimide (PI). Finally, we discuss the range of conductances which we expect can be measured with this technique.

\section{Measurement Principle}
\begin{figure}
\includegraphics{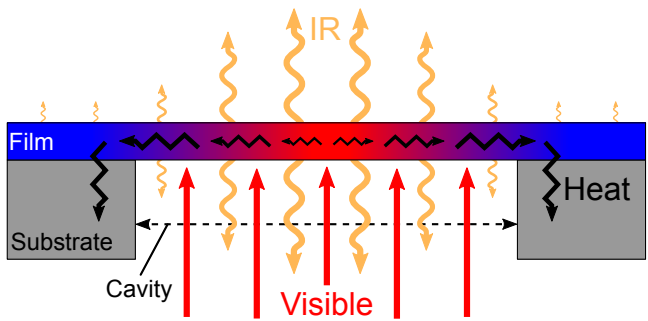}
\caption{\label{fig:method} A film suspended over a hole in an opaque substrate. Illumination in the visible range (straight red arrows) from below heats the free-standing film homogeneously. Heat (jagged black arrows) is transferred along the film into the substrate acting as a heat sink. In steady state a maximum temperature of the free-standing film and a maximum thermal radiation (wavelike orange arrows) is reached in the center of the hole.}
\end{figure}

In order to measure a thermal conductivity, a temperature gradient needs to be generated by a known heat flux. Figure~\ref{fig:method} shows a schematic diagram of the method used here to generate such a temperature gradient, where thin films are placed over a cavity in a substrate. The sample is illuminated from below, by which the required known heat flux is generated. The opaque substrate serves as a shadow mask for the illumination and as a heat sink, generating the necessary temperature gradient. The resulting temperature distribution of the free-standing part of the film depends on the form of the cavity. Using analytical or numerical models the thermal conductivity can then be obtained from the magnitude of the temperature increase taking into account the illumination power density, the absorptance of the film, and its thickness. These are the only quantities that need to be measured in addition to the temperature distribution for the determination of the thermal conductivity via this method.

In this work we concentrate on circular cavities. For free-standing thin films suspended over circular cavities with radius $R$ (and also assuming a homogeneous absorption of the illumination over the depth of the film), rotational invariance allows an analytical solution of the temperature distribution. In cylindrical coordinates the resulting one-dimensional problem can be written as
\begin{equation}
q_r (r)=-\kappa_{||}  \frac{\partial T(r)}{\partial r}
\label{eq:fourier}
\end{equation}
with the radial component of the heat flux vector $q_r$, the in-plane thermal conductivity $\kappa_{||}$, and the temperature distribution $T(r)$ at a distance $r$ from the cavity center.\cite{Ser14} For a homogeneously absorbed areal power density $p_\mathrm{abs}$, symmetry leads to a vanishing heat flux at the center of the hole ($r=0$). At steady-state conditions the power absorbed by a circle area $r^2 \pi$ needs to be equal to the power flowing through the circumference of said circle. This leads to
\begin{equation}
r^2 \pi \cdot p_\mathrm{abs}=2 r \pi \cdot d \cdot q_r (r)
\label{eq:EnergyConservation}
\end{equation}
where $d$ is the film thickness.

Merging (\ref{eq:fourier}) and (\ref{eq:EnergyConservation}), integrating, and applying the boun\-dary condition $T(r=R)=T_\mathrm{substrate}$, i.e.~the substrate temperature, leads to the temperature distribution
\begin{equation}
T(r)=-\frac{p_\mathrm{abs}}{4\kappa_{||} d}r^2+\frac{p_\mathrm{abs}}{4\kappa_{||} d}R^2+T_\mathrm{substrate},
\label{eq:TDistribution}
\end{equation}
expected for the experimental geometry investigated here.

The simple boundary condition used does not take into account the exact transition from the parabolic behavior in the middle of the free-standing film to the substrate-supported film. However, this transition, influenced by thermal contact resistance and shadowing at the border of the hole, will also be rotationally invariant leading to an additional constant offset to the temperature distribution. This offset and the constant term $\frac{p_\mathrm{abs}}{4\kappa_{||} d}R^2+T_\mathrm{substrate}$ in (\ref{eq:TDistribution}) can be incorporated into an effective substrate temperature $T_\mathrm{substrate, eff}$. Therefore, $\kappa_{||}$ can be directly determined from the curvature of the temperature distribution near the center of the hole, $p_\mathrm{abs}$, and $d$ only. Since the curvature can be obtained from a measurement of local temperature differences the demand for the absolute temperature accuracy is minimal. 

Furthermore, the above derivation of (\ref{eq:TDistribution}) assumes vanishing losses, e.g. from convection and thermal radiation, generating additional cooling processes. We will discuss thermal radiation losses towards the end of the paper. Losses through air are reduced by a vacuum chamber, described below.

\section{Setup details}
\label{sec:setup}
\begin{figure}
\includegraphics{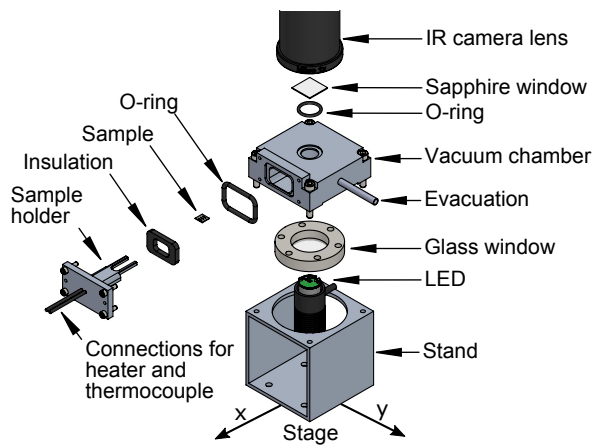}
\caption{\label{fig:setup} Exploded view of the measurement setup. IR thermography of the sample in vacuum is performed from above through a sapphire window. The sample holder allows to heat the sample and to measure its temperature directly via thermocouples. The heating of the thin film is achieved optically via illumination from below through a glass window. A translation stage allows lateral positioning.}
\end{figure}

Integral parts of the setup are the IR camera for the spatially resolved temperature measurement and the vacuum chamber, shown in Fig.~\ref{fig:setup}, for the reduction of the losses by air.
The IR camera used was an InSb 640 SM high speed camera from DCG systems equipped with a Micro 1x lens, that projects the sample image with a 1:1 scale onto the detector of the camera. The resulting distance on the sample between two pixels in the camera image $l$ was determined to be \SI{14.7}{\micro\meter} and the resolution of the camera is 640x480 pixels, giving a total field of measurement of \SI{9.4x7.1}{\milli\meter}. The noise equivalent differential temperature (NEDT) is below \SI{20}{\milli\kelvin} and for each measurement at least 128 images were averaged at a frame rate of \SI{300}{\hertz}.

The IR camera observes the sample from above. For the window material to the vacuum chamber we use sapphire to allow the transmission of the IR radiation. The sample consists of sheets of aluminum with a thickness of \SI{0.3}{\milli\meter} and holes ranging from \SI{0.3}{\milli\meter} to \SI{0.8}{\milli\meter} in diameter as the opaque substrate with the film under investigation on top. The aluminum substrate allows simple manufacturing and offers a large thermal conductivity for fast removal of excess heat. It also reflects most of the incident light, thus the increase of the substrate temperature was always below \SI{1}{\kelvin}, even for the highest illumination powers.

The thickness of the films was determined by placing the films on a glass substrate and using profilometry. The results were confirmed with scanning electron microscopy and atomic force microscopy. Since the thickness $d$ of the films investigated here is orders of magnitude below the hole radius $R$, the slight inhomogeneity in absorption with respect to the depth in the film can be neglected. 

The whole sample is suspended underneath the sapphire window and is illuminated from below via an LED. The LED is placed outside of the vacuum chamber and irradiates the sample through a glass window, which blocks the IR radiation of the hot LED due to its low IR transmittance. The visible light of the LED itself is not detected by the IR camera. The sample holder has an integrated heater and a feedthrough for a thermocouple in contact with the substrate. Heater and thermocouple are used for the calibration summarized below. In order to reduce background radiation, thermal insulation assures that the chamber temperature remains constant while the sample and holder are heated. The setup is mounted on a translation stage for sample positioning and measurement of the illumination power density.

\begin{figure}
\includegraphics{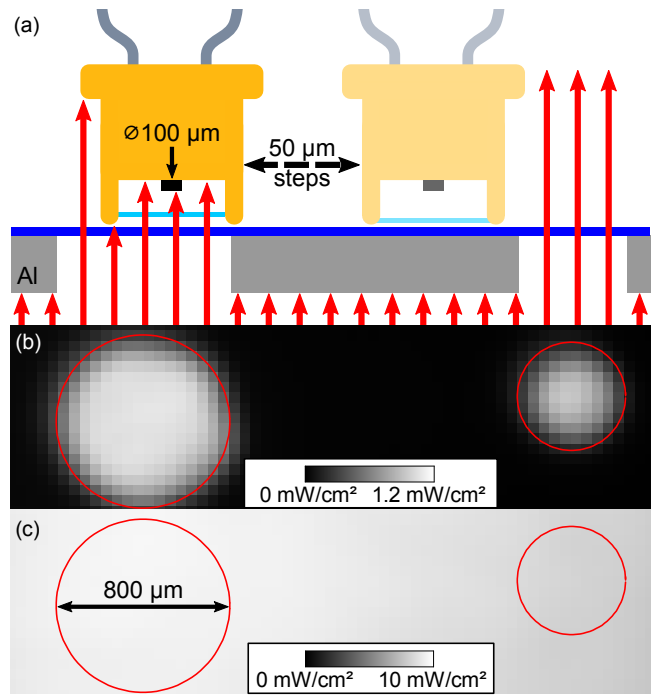}
\caption{\label{fig:Scan} (a) Scanning the sample with a calibrated photo-diode (black) with a diameter of \SI{100}{\micro\meter} by moving the chamber and a blue LED underneath the fixed diode by \SI{50}{\micro\meter} steps. (b) and (c) show the measured power density with and without a PI sample on a supporting substrate, respectively. The red circles illustrate the position of the holes. The difference in absolute intensity between (b) and (c) is due to the low transmittance of PI at the used illumination wavelength.}
\end{figure}

For the latter, a circular silicon photo-diode with a diameter of \SI{100}{\micro\meter} is fixed above the film, pointing in the direction of the LED, as shown in Fig.~\ref{fig:Scan}(a). The photo-diode was cross-calibrated with a NIST-traceable power sensor. The chamber with the LED and the sample are then scanned underneath the photo-diode, yielding the map of the holes in Fig.~\ref{fig:Scan}(b). In a second step the sample, consisting of the film and the supporting substrate, is removed and the scan repeated for all LED intensities used for the thermal conductivity measurement. This allows the acquisition of the spatially resolved illumination power density in Fig.~\ref{fig:Scan}(c). This power density includes the effects of the glass window and allows verification of the lateral homogeneity. The hole position from Fig.~\ref{fig:Scan}(b) is than correlated to the power density in Fig.~\ref{fig:Scan}(c) and the power density for each hole is determined by averaging the values within each red circle.

The absorptance $A$ in the range of the optical wavelengths used for heating and the IR transmittance and reflectance of the films are determined via UV-Vis and FTIR transmission and reflection measurements, respectively. For UV-Vis spectroscopy an integrating sphere was used to include diffuse scattering. For these experiments the thin films were again placed on sheets of aluminum with a hole. This time the holes had a diameter of \SI{10}{\milli\meter} in order to increase the signal intensities. 

Once the absorptance $A$ of the film under investigation as well as the illumination power density is known the areal power density $p_{abs}$ can be calculated providing the desired known heat flux.

\begin{figure}
\includegraphics{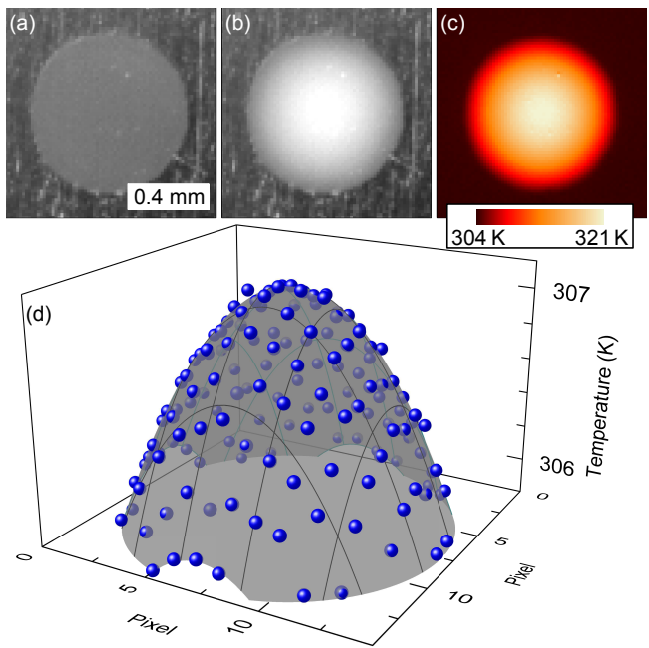}
\caption{\label{fig:TempDisp} (a) and (b) IR images of a \mbox{PEDOT:PSS} film over a cavity with a diameter of \SI{0.8}{\milli\meter} in the dark and under illumination, respectively. (c) False color temperature map derived from (a) and (b) after calibration. (d) Temperature distribution of the same film over a hole with a diameter of \SI{0.3}{\milli\meter} under illumination (blue dots). The gray surface shows a fitted paraboloid. The sample in images (b) through (d) was illuminated with a power density of \SI{42}{\milli\watt\per\centi\meter^2}.}
\end{figure}

\section{Temperature Measurement}
We calibrate the IR camera, the second important step in the method presented here, directly with the films investigated, and thereby also account for the emissivity of the film.

The signal detected by the IR camera is proportional to the IR emission of the film in the detection wavelength range of the camera. For the temperature calibration of the camera two images of the sample at different temperatures, as determined by the thermocouple, are taken. An example of such a calibration image without illumination can be seen in Fig.~\ref{fig:TempDisp}(a). For small temperature differences the calibration can be approximated by a linear relation between the temperature and the camera signal. More accurate results, however, are obtained when the nonlinear dependence of the number of photons $N_\mathrm{photon}$, emitted by the film at a temperature $T$, is taken into account. This correlation is given by the photon radiance
\begin{equation}
N_\mathrm{photon}(\lambda) \propto \frac{1}{\lambda^4 \left[\exp{\left(\frac{hc}{\lambda k_\mathrm{B} T}\right)}-1 \right]}~,
\label{eq:planck}
\end{equation}
where $\lambda$ is the wavelength, $k_\mathrm{B}$ the Boltzmann constant, $h$ the Planck constant, and $c$ the speed of light. This dependence is treated numerically. The dependences of the lens, the camera, and the film emissivity on $\lambda$ are effectively removed by using a narrow band IR filter in front of the camera that is transparent between \SI{4.5}{\micro\meter} and \SI{5}{\micro\meter} only. While the application of this correction in principle requires knowledge of the absolute temperature, simulations have shown that the effect of the exact absolute temperature on the $\kappa_{||}$ determined is minimal.

After calibration, an IR image of the film under illumination can be taken, shown in Fig.~\ref{fig:TempDisp}(b). From this the temperature distribution in Fig.~\ref{fig:TempDisp}(c) can be calculated, including the correction for the slight non-linearity in (\ref{eq:planck}), which is then used to deduce the in-plane thermal conductivity. For fitting, only the center of the film is used, as discussed above in conjunction with (\ref{eq:TDistribution}). In Fig.~\ref{fig:TempDisp}(d) the central temperature distribution of the same film over a smaller hole is compared to a paraboloid fitted to this data. Clearly, the temperature distribution in blue is described well by the first term in (\ref{eq:TDistribution}), indicated by the gray surface, which allows the direct determination of $\frac{p_\mathrm{abs}}{4 \kappa_{||} d}$ and, together with the experiments described in Sec.~\ref{sec:setup}, the measurement of $\kappa_{||}$.

\section{Sample Preparation}
The method presented here requires a sample that has a sufficient emissivity in the detection range of the camera and a sufficient absorptance at the wavelength of the illumination used for heating. For this work we have studied two different thin film materials that comply with these requirements, \mbox{PEDOT:PSS} and PI. While \mbox{PEDOT:PSS} is an example of a novel organic semiconductor, PI is a rather standardized material, but nonetheless offers an interesting study case as varying values for the thermal conductivity $\kappa_{||}$ are found in the literature, which we will discuss later.

Free-floating thin films of PEDOT:PSS on water were produced by the method described in Ref.~\onlinecite{Gre11}. The \mbox{PEDOT:PSS} solution, containing \SI{1}{\percent} Zonyl, was spin-coated thrice onto a glass substrate covered with a thin layer of poly(dimethylsiloxane) (PDMS). Such a thin layer of PDMS reduces adhesion of the \mbox{PEDOT:PSS} film to the substrate, allowing for easier lift-off later. After each spin coating the samples were annealed for \SI{10}{\minute} at \SI{140}{\celsius}. Post-treatment was carried out by drop-casting ethylene glycol (EG), letting it soak in for \SI{1}{\minute}, spinning it off, and annealing again for \SI{10}{\minute} at \SI{140}{\celsius}. This yielded an electrical conductivity $\sigma=\SI{720\pm30}{\siemens\per\centi\meter}$ at a typical film thickness of \SI{250}{\nano\meter}.\cite{Pal14} Poly(vinylalcohol) (PVA) was drop-cast onto the sample and after drying, the PVA/\mbox{PEDOT:PSS} film could easily be removed from the PDMS surface. The PVA was then dissolved in water and the remaining free-floating \mbox{PEDOT:PSS} film was transferred from the water onto the desired substrate. The adhesion of the PEDOT:PSS to the final substrate was very good, ensuring sufficient thermal contact with the support.

We also investigated commercial type HN polyimide films with a thickness of \SI{9}{\micro\meter}. Transfer was much simpler for the PI, however, the adhesion and thus the thermal contact to the substrate were inferior. To improve adhesion and thermal contact, thermal grease was applied between the aluminum and the PI film. 

\section{Results and Discussion}
\begin{figure}
\includegraphics{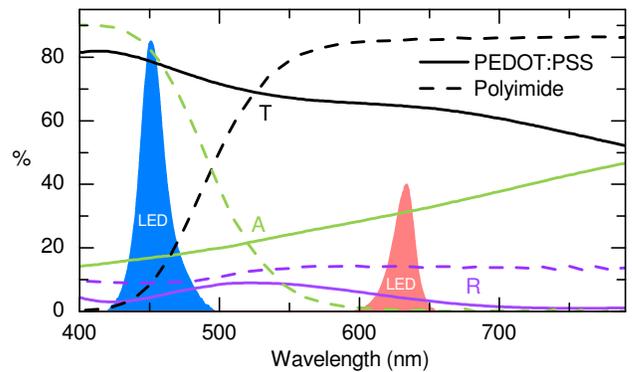}
\caption{\label{fig:UVVis} UV-Vis measurements of the investigated PI and \mbox{PEDOT:PSS} films, shown as dashed and solid lines, respectively. An integrating sphere was used to measure the transmittance (black) and reflectance (violet) of the films and the absorptance (green) was calculated with $A+T+R=1$. The spectral data of the blue and red LED are shown in arbitrary units to indicate the absorptance of the PI and PEDOT:PSS films, respectively.}
\end{figure}

We start with the optical characterization of the thin films in the visible spectral range used for heating. The transmittance $T$ and reflectance $R$ are plotted in Fig.~\ref{fig:UVVis} for the \mbox{PEDOT:PSS} and PI films. The absorptance $A$ for each film was then calculated according to $A+T+R=1$. For direct comparison, the spectra emitted by the two LEDs used with mean wavelengths of \SI{632}{\nano\meter} and \SI{455}{\nano\meter} are also plotted in Fig.~\ref{fig:UVVis} in arbitrary units in red and blue, respectively. The highest absorptance was achieved by using the red LED for the \mbox{PEDOT:PSS} film and the blue LED for the PI film. Weighting the absorptance spectra with the LED spectra results in effective absorptances of \SI{0.3}{} and \SI{0.8}{}, respectively. 

Additionally, the IR  transmittance of the films was measured via FTIR spectroscopy to determine the influence of the increase in background radiation during calibration, caused by an unavoidable slight increase of the chamber temperature. For the \mbox{PEDOT:PSS} film, the IR transmittance is below \SI{10}{\percent} while the increase in background radiation during calibration is less than \SI{20}{\percent} of the increase in film radiation from holes covered by the film. Therefore the total background influence on the calibration is below \SI{2}{\percent} and can be neglected. The IR transmittance of the PI film studied is significantly higher, ranging between \SI{70}{\percent} and \SI{90}{\percent}. At the same time the emissivity of the PI film is significantly lower than that of the PEDOT:PSS film and thus the increase in background radiation is about \SI{40}{\percent} of the increase in film radiation. This is corrected for in the calibration.

\begin{figure}
\includegraphics{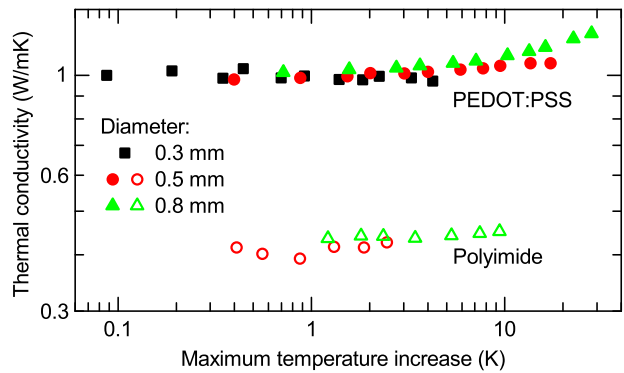}
\caption{\label{fig:cond} The in-plane thermal conductivity of \mbox{PEDOT:PSS} films with a thickness of \SI{250}{\nano\meter} and PI films with a thickness of \SI{9}{\micro\meter} as a function of the maximum temperature increase resulting from the LED illumination. The illumination power density was varied between \SI{2}{\milli\watt\per\centi\meter^2} and \SI{100}{\milli\watt\per\centi\meter^2}.}
\end{figure}

Together with the film thicknesses, we now have all data available to determine $\kappa_{||}$ via steady-state thermography. The results for $\kappa_{||}$ are shown in Fig.~\ref{fig:cond} as a function of the maximum temperature increase, measuring films on holes with different diameters and with different illumination power densities. As expected, a larger hole diameter and a higher illumination power lead to a larger temperature increase. For an increase below \SI{4}{\kelvin} the results for \mbox{PEDOT:PSS} show a stable value of $\kappa_{||}=\SI{1.00\pm0.04}{\watt\per\meter\per\kelvin}$ for all hole diameters with the peak-to-peak distribution of values given in parenthesis. For a higher temperature increase the measured values of $\kappa_{||}$ increase. This is most likely due to increased thermal radiation losses already mentioned earlier, proportional to $T^4$, which were not corrected. 

For PI we obtain $\kappa_{||}=\SI{0.42\pm0.02}{\watt\per\meter\per\kelvin}$. Because of a larger film thickness the PI does not heat up as much as the PEDOT:PSS for the same power densities. Thus the radiation losses are smaller and the apparent increase in $\kappa_{||}$ for large temperature increases is less pronounced. 

We now estimate the uncertainty in the determination of $\kappa_{||}$. The fit error and the influence of the camera noise depend on the maximum temperature increase. For increases above \SI{2}{\kelvin}, they are typically below \SI{1}{\percent} and can be neglected. Uncertainties for the absorptance measured by UV-Vis spectroscopy, for the thickness determination, and for the illumination power were all estimated at below \SI{10}{\percent} and uncertainties for the temperatures determined by the thermocouples at below \SI{5}{\percent}. For the PI film measurements an additional error of \SI{10}{\percent} was estimated, accounting for the calibration correction due to background radiation.

The influence of thermal radiation losses of the thin films on the calibration and the actual measurements is more challenging to estimate. Simulations have shown that radiation losses during calibration and LED illumination cancel each other out for small illumination powers and hole diameters. This is finally supported by the data in Fig.~\ref{fig:cond}, where the values for $\kappa_{||}$ are independent of the hole diameter for low illumination powers. We conservatively account for the effect of radiation losses with another \SI{10}{\percent} uncertainty. This leads to values for $\kappa_{||}$ of \SI{1.0\pm0.2}{\watt\per\meter\per\kelvin} and \SI{0.42\pm0.09}{\watt\per\meter\per\kelvin} for the investigated PEDOT:PSS and PI films, respectively, with the errors now accounting for all parts of the experimental method.

Our results are in very good agreement with literature values of the in-plane thermal conductivity of \mbox{PEDOT:PSS} after post-treatment. The results for dimethyl sulfoxide- (DMSO) and EG-doped \mbox{PEDOT:PSS} in Ref.~\onlinecite{Liu15} and Ref.~\onlinecite{Wei14}, respectively, are within our uncertainties. Those measurements were also performed on PEDOT:PSS that was prepared on PDMS and eventually removed from the substrate, as was done here.\cite{Gre11} Reference~\onlinecite{Kim13} and Ref.~\onlinecite{Zha10} reported lower values for $\kappa_{||}$, however, these samples were not prepared on PDMS and the results were deduced from cross-plane thermal conductivities. 

As mentioned above varying values for the thermal conductivity $\kappa$ of PI can be found depending on PI type, film thickness, and measurement orientation.\cite{Ben99,Yok95,Cho87} The manufacturer of the PI film studied here, DuPont, and the reseller, GoodFellow, however, state a thermal conductivity $\kappa$ of \SI{0.12}{\watt\per\meter\per\kelvin} and \SI{0.16}{\watt\per\meter\per\kelvin}, respectively, for bulk and thin films alike. The measurement geometry used to determine $\kappa$ is not given. A more in-depth study by Rule~\textit{et~al.}, using an unguarded steady-state parallel-plate apparatus and stacking 33 layers of PI film, found a thermal conductivity of $\kappa_{||}=\SI{0.38}{\watt\per\meter\per\kelvin}$ for the in-plane thermal conductivity of thin films of type HN PI, the type of PI studied here, again well within our uncertainties.\cite{Rul96}

\begin{figure}
\includegraphics{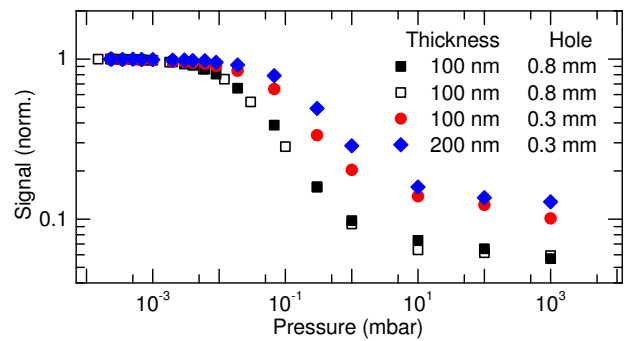}
\caption{\label{fig:vac} Normalized maximum increase of the thermography signal of optically heated PEDOT:PSS films for varying environmental pressure. Solid symbols and open symbols represent illumination power densities of \SI{7}{\milli\watt\per\centi\meter^2} and \SI{22}{\milli\watt\per\centi\meter^2}, respectively. Thinner films and larger hole diameters show a stronger pressure dependence.}
\end{figure}

Finally we briefly address the necessity to perform these measurements in vacuum. Figure~\ref{fig:vac} shows the dependence of the maximum temperature increase of illuminated PEDOT:PSS films on the chamber pressure.
It becomes clear that the signal detected is significantly lower at atmospheric pressure, where the surrounding gas acts as an additional heat sink. The thermal conductivity of the air then follows the characteristics of the Pirani pressure gauge\cite{Woo61,Shi95} before saturating for a pressure below \SI{1e-3}{\milli\bar} where the influence of the remaining gas can be neglected. Worth mentioning is the independence of this effect on the LED power, evident from comparing the open and solid black squares in Fig.~\ref{fig:vac}, as well as the fact that thicker films show less influence because of a reduced surface-to-volume ratio.

\section{Range of thermal conductivities accessible}

Finally we want to estimate the range of thermal conductivities which might be measured with the method presented here. Large thermal conductivities will result in a small temperature increase. To be able to detect this, high optical heating powers and large cavity diameters will be necessary. To estimate the maximum thermal conductivity accessible, we conservatively assume that the minimum signal that can be achieved by the highest LED power over the largest hole diameter should be equivalent to the maximum temperature increase of \SI{1}{\kelvin} for our \mbox{PEDOT:PSS} film, which has an emissivity of  $\epsilon_\mathrm{ir}=0.3$ as estimated by FTIR. This would still allow a measurement series with varying illumination power and hole diameters to confirm the independence of the value of $\kappa_{||}$ on power and diameter as in Fig.~\ref{fig:cond}, while offering temperature differences well above the NETD of the IR camera. When further assuming an ideal emissivity of $\epsilon_\mathrm{ir}=1$ of the film investigated, a temperature increase $\Delta T=\SI{0.3}{\kelvin}$ should be detectable. Equation (\ref{eq:TDistribution}) can now be used to link this temperature increase, the hole radius $R$, the optical heating power density $p_\mathrm{abs}$, and the film thickness $d$ via 
\begin{equation}
\Delta T=-\frac{p_\mathrm{abs}}{4\kappa_{||} d}R^2.
\label{eq:tempinc}
\end{equation}
The largest hole radius that can be captured by the camera using the 1x microscope objective with the 1:1 image scaling is $R\approx\SI{5}{\milli\meter}$. With an illumination power density of \SI{100}{\milli\watt\per\centi\meter^2} this leads to the maximum thermal conductivity detectable of
\begin{equation}
\kappa_\mathrm{max}=\frac{\epsilon_\mathrm{ir}A}{d} \cdot \SI{0.02}{\watt\per\kelvin},
\label{eq:kappamax}
\end{equation}
where A is the absorptance of the thin film.

For low thermal conductivities radiation losses are the limiting factor. The radiation losses per area for both sides of the film can be expressed as 
\begin{equation}
p_\mathrm{rad}=2 \epsilon_\mathrm{tot}\sigma\left(\left(\Delta T+T_\mathrm{amb}\right)^4-T_\mathrm{amb}^4\right),
\label{eq:radloss}
\end{equation}
with the total emissivity of the film $\epsilon_\mathrm{tot}$, the Stefan-Boltzmann constant $\sigma$, and the ambient temperature $T_\mathrm{amb}$. First order approximation and division by $p_\mathrm{abs}$ from (\ref{eq:tempinc}) leads to
\begin{equation}
\frac{p_\mathrm{rad}}{p_\mathrm{abs}}=2 R^2 \frac{\epsilon_\mathrm{tot}\sigma T_\mathrm{amb}^3}{\kappa_{||} d}.
\label{eq:radfigure}
\end{equation}
This can be used as a figure of merit for the influence of the radiation losses on the measurement. As an upper limit for this figure we choose the measurement of \mbox{PEDOT:PSS} over a hole with a diameter of \SI{0.5}{\milli\meter}, since for low illumination power densities again no dependence of $\kappa_{||}$ on the variation of the experimental conditions is observed. With a total emissivity of the film of  $\epsilon_\mathrm{ir}=0.3$, again estimated by FTIR, the ratio (\ref{eq:radfigure}) should thus be below 0.2. A simple way to realize such a value of $\frac{p_\mathrm{rad}}{p_\mathrm{abs}}$ is the reduction of $R$, however, this is limited by the resolution of the IR camera. Assuming that 15 pixels should be the minimum for the image of a hole diameter to assure a satisfying fit (cf. Fig.~\ref{fig:TempDisp}), we find again a minimum $R$ of \SI{0.11}{\milli\meter}, using the 1x microscope objective. With (\ref{eq:radfigure}) the minimum thermal conductivity detectable by this setup is thus
\begin{equation}
\kappa_\mathrm{min}=\frac{\epsilon_\mathrm{tot}}{d} \cdot \SI{2e-7}{\watt\per\kelvin}.
\label{eq:kappamin}
\end{equation}

The limits for $\kappa_{||} \cdot d$, that we expect should be detectable with the setup developed here, can thus be written as
\begin{equation}
\epsilon_\mathrm{ir}A \cdot \SI{0.02}{\watt\per\kelvin} \gtrsim \kappa_{||}d \gtrsim \epsilon_\mathrm{tot} \cdot \SI{2e-7}{\watt\per\kelvin}.
\label{eq:limitsetup}
\end{equation}
This means, e.g, that a black sheet of aluminum with a thermal conductivity of \SI{237}{\watt\per\meter\per\kelvin} and assuming $A=1$ and $\epsilon_\mathrm{ir}=1$ should have a maximum thickness of \SI{85}{\micro\meter} to be within this thermal conductivity range.\cite{Ho72} A \mbox{PEDOT:PSS} film on the other hand should have a minimum thickness of \SI{60}{\nano\meter}, assuming that the optical properties remain as they were observed in this work.

Using a 10x microscope objective and thereby reducing the distance on the sample between two pixels in the camera image to $l=\SI{1.5}{\micro\meter}$ would potentially reduce the lower limit by a factor of 100, but at the cost of more noise.
Generalizing our findings to a different camera setup we get 
\begin{equation}
0.016 \cdot \frac{\epsilon_\mathrm{ir} p_\mathrm{max} A R_\mathrm{max}^2}{NEDT} \gtrsim \kappa_{||}d \gtrsim \epsilon_\mathrm{tot} l \cdot \SI{0.013}{\watt\per\meter\per\kelvin},
\label{eq:limitany}
\end{equation}
with the maximum illumination power density $p_\mathrm{max}$ and the maximum possible radius $R_\mathrm{max}$.

\section{Conclusion}
In summary, we have presented a fast contact-free method for the determination of the in-plane thermal conductivity of free-standing thin films and outlined the method's possible range of application. After calibration the measurement of samples with sizes between \SI{10}{\milli\meter} and \SI{10}{\micro\meter}, depending on the microscope objective, with multiple measurement cavities can be performed in a few seconds, providing statistically relevant results and allowing to judge the homogeneity of the samples investigated. The results obtained on the low conductivity test materials used agree quantitatively with the existing literature. With on-sample calibration, no additional knowledge for the temperature calibration is necessary, thus the technique could be applied to novel materials such as highly porous SiGe films, as well as composite and organic-inorganic hybrid materials.\cite{Sto15,Sch16,Car12,See10b} The setup presented here furthermore allows to measure $\kappa_{||}$ as a function of temperature (by additionally heating the substrate in the vacuum chamber) and the application to brittle materials (due to the small hole diameters possible). While for large temperature differences generated by the optical heating losses due to thermal radiation were identified as an issue, for low temperature differences consistent results were observed. Should this requirement lead to signal amplitudes too small to be measured accurately with the steady-state thermography presented here, the application of quasi-steady-state lock-in methods could be helpful. Eventually leaving the steady-state regime, an investigation of the heating dynamics should even yield the heat capacity of the film under investigation with the measurement geometry presented here. 

\section{Acknowledgments}
We gratefully acknowledge funding by the Bavarian State Ministry of the Environment and Consumer Protection via the project UMWELTnanoTECH, by the Bavarian State Ministry of Education, Science, and the Arts via a project from Energy Valley Bavaria at the Munich School of Engineering, and by the Nanosystems Initiative Munich (NIM).

\hypersetup{hidelinks}
\bibliography{ThermalCond_AntonGreppmair}

\end{document}